\documentclass[showpacs,twocolumn]{revtex4}
\usepackage{amsthm}   
\usepackage{amsmath}  
\usepackage{amssymb}  
\usepackage{mathrsfs}
\usepackage[all]{xy}
\usepackage[pdftex]{graphicx}
\usepackage{indentfirst}
\usepackage[pdftex]{hyperref}

\newcommand{\be}{\begin{equation}}
\newcommand{\ee}{\end{equation}}
\newcommand{\ba}{\begin{eqnarray}}
\newcommand{\ea}{\end{eqnarray}}

\newcommand{\en}{\nonumber\\}

\newcommand{\de}{\delta}

\newcommand{\dd}[1]{\dot{#1}}

\newcommand{\kk}{\mathbf{k}}

\newcommand{\kkk}{\mathbf{k}'}

\begin{document}
\title{Study of the Growth of Entropy Modes in MSSM Flat Directions Decay:\\ Constraints on the Parameter Space}
\author{Francis-Yan Cyr-Racine}
\affiliation{Department of Physics and Astronomy, University of British Columbia,6224 Agricultural Road, Vancouver, BC, V6T 1Z1, Canada}
\author{Robert H. Brandenberger}
\affiliation{Department of Physics, McGill University, 3600 University Street, Montreal, QC, H3A 2T8, Canada \\ and \\
Institute of High Energy Physics, Chinese Academy of Sciences, P.O. Box 918-4, 
Beijing 100049, P.R. China}
\date{\today}


\begin{abstract}

We study how the resonant decay of moduli fields arising in the Minimal Supersymmetric Standard Model (MSSM) could affect large scale curvature perturbations in the early universe. It has been known for some time that the presence of entropy perturbations in a multi-component system can act as seeds for the curvature perturbations on all scales. These entropy perturbations could be amplified exponentially if one of the moduli decays via stochastic resonance, affecting the curvature power spectrum in the process. By imposing the COBE normalization on this power spectrum, one could put constraints on the masses and couplings of the underlying particle physics model without having to rely on collider experiments. We discuss in detail the case of the MSSM but this method could be applied to other theories beyond the Standard Model.

\end{abstract}

\maketitle

\section{Introduction}

Many models of physics beyond the Standard Model contain light scalar fields \cite{Iliopoulos:2008fc,Allanach:2005fm} which are usually called moduli fields. These field could have played an important role in the early Universe and therefore cosmology is a natural test-bed to study the properties of the moduli. Generically, models beyond the Standard Model predict the existence of \emph{multiple} moduli fields. In such multi-component systems, one must consider the presence of entropy perturbations between the different constituents of the system in the early universe (see e.g. \cite{Gordon:2000hv}). 

It is well-known that entropy modes (associated with non-adiabatic pressure perturbations) can act as a source for curvature perturbations on all scales (\cite{Liddle:1999hq} and references therein). Generally, the contribution of the entropy modes to the curvature power spectrum is small and can be neglected. 
However, under certain circumstances, the non-perturbative decay of one modulus field (e.g. the inflaton itself) can exponentially amplify the amplitude of an entropy perturbation mode \cite{BV,FB2,Zibin,Liddle:1999hq}. If this happens, the contribution to the curvature power spectrum arising from entropy perturbations could dominate over the adiabatic modes (as in the curvaton scenario, see e.g. \cite{McDonald:2003jk}). By demanding that the curvature power spectrum respect the bound set by the COBE normalization \cite{Bunn:1996py}, one can put constraints on moduli parameters (masses and couplings), hence constraining certain models of physics beyond the Standard Model (see \cite{Larissa,Keshav} for applications of this idea to string inflation models).

The Minimal Supersymmetric Standard Model (MSSM) is a good example of a model having a large moduli space \cite{Enqvist:2003gh}. Indeed, the scalar potential of the MSSM has a large number of flat directions along which the potential vanishes \cite{Gherghetta:1995dv}. Supergravity corrections usually spoil the flatness of the potential \cite{Allahverdi:2001is}. Nevertheless, there exists a large class of models with non-minimal Kahler potential for which scalar fields can develop large vacuum expectation values (vevs) along flat directions during inflation \cite{Gaillard:1995az,Kasuya:2006wf}. 

After inflation, the vev closely tracks the minimum of the potential as it evolves to lower scales \cite{Dine:1995kz}. Once the Hubble parameter is of the order of the soft SUSY-breaking scale, the flat direction starts oscillating about its true minimum. Assuming an F-term interaction of the form $\Delta V=g^2|\varphi|^2|\chi|^2$, where $\varphi$ is the moving vev and $\chi$ is some MSSM field, these oscillations could trigger the non-perturbative production of $\chi$ particles whenever the change in 
the frequency of $\chi$ ceases to be adiabatic ($\dd{\omega}_{\chi}/\omega_{\chi}^2>1$) \cite{Kofman:1997yn}.

However, since flat directions are inherently complex fields, they do not typically pass through $|\varphi|=0$ during their oscillations, ensuring that $\dd{\omega}_{\chi}/\omega_{\chi}^2\ll1$ at all times \cite{Postma:2003gc}. Nevertheless, under certain circumstances, the ellipticity of the trajectory in field space can be small enough such that one can apply the well-know results from parametric resonance \cite{TB}.

Furthermore, it was recently shown \cite{Olive:2006uw,Gumrukcuoglu:2008fk,Basboll:2007vt,Basboll:2008gc} that if two or more flat directions are simultaneously excited, the rotation of the vevs  could trigger exponential production of MSSM fields. Thus, even if the ellipticity of the trajectory is large, we expect entropy modes between the flat direction condensates and the created particles to be amplified.

In this paper, we focus on the simpler case where the motion of the oscillating vev is purely radial. We consider the case of a quantum field $\chi$ whose mass during inflation is somewhat smaller than the Hubble scale and which is coupled to a flat direction.  During inflation, quantum fluctuations of this field are stretched to super-Hubble scales and form an almost scale-invariant spectrum at the end of inflation. The subsequent decay of the flat direction triggers an exponential amplification of these super-Hubble modes via stochastic resonance \cite{Kofman:1997yn}. These modes correspond to an entropy perturbation that then acts as a source for curvature perturbations. We compute the curvature power spectrum and compare it to the bound set by the COBE normalization.  Note that induced curvature fluctuations from entropy modes in supersymmetric models
were recently also studied in \cite{Riotto}. New to our work is the focus on the parametric amplification of the entropy
mode during the phase of oscillation of a flat direction field.


\section{The Model}

We consider a MSSM flat direction parametrized by the order parameter $\varphi$. Soft SUSY-breaking and Hubble-induced mass terms as well as superpotential corrections of the type $W\supset\lambda\varphi^n/nM^{n-3}$ ($n>3$) lift the flat direction by giving it a potential of the form \cite{Dine:1995kz}:
\be\label{pot_final}
V(\varphi) \, = \, (m_{\varphi}^2-c_HH^2)|\varphi|^2+|\lambda|^2\frac{|\varphi|^{2n-2}}{M^{2n-6}},
\ee
where $M$ is the scale at which the non-renormalizable terms become important. Here we neglected soft SUSY-breaking and Hubble-induced A-terms. With a minimal choice of Kahler potential, $c_H=-3$ and the flat direction mass is positive. Therefore, $\varphi$  settles at the origin during inflation and does not lead to any interesting cosmological consequences. Nevertheless, there exists a large class of Kahler potentials for which $c_H>0$ \cite{Gaillard:1995az,Kasuya:2006wf,Allahverdi:2007zz}. In these cases, the potential (\ref{pot_final}) admits a minimum at:
\be\label{varphi_inf}
|\varphi|_{min} \, = \, \left(\frac{2\sqrt{(n-1)c_H} HM_p^{n-3}}{|\lambda|}\right)^{\frac{1}{n-2}},
\ee
where we took the cutoff to be the Planck scale, $M=M_p$. During inflation, the flat direction settles down at this minimum. The renormalizable MSSM superpotential \cite{Gherghetta:1995dv} naturally provides couplings between the modulus field $\varphi$ and other MSSM fields $\chi_i$. The generic form of the potential for a field coupled to flat direction via an F-term interaction is:
\be
V(\chi) \, = \, \frac{1}{2}m_{\chi}^2\chi^2+\frac{1}{2}g^2|\varphi|^2\chi^2.
\ee 
In the following, we will always assume that $m_{\chi}\ll g|\varphi|$ such that we can neglect the mass of the $\chi$ field. 

During inflation, we demand that 
$g^2|\varphi_{min}|^2 <  H^2_I $
in order for $\chi$ to acquire un-suppressed fluctuations on large scales. This can be achieved by choosing the smallest possible value of $n$, namely $n = 4$. Indeed, $|\varphi|_{min}$ lies parametrically between $H$ and $M_p$ and it is easy to see that for large $n$, $|\varphi|_{min}\rightarrow M_p$. A way to relax this condition would be to consider $\sqrt{2c_H}/\lambda\ll 1$. However, we need $c_H\sim\mathcal{O}(1)$ \cite{Dine:1995kz} in order for the quantum fluctuations of $\chi$ not to dominate the mass of $\varphi$ (otherwise $\varphi$ would not oscillate freely).  This in turn implies $\lambda\gg1$ in order to have $\sqrt{2c_H}/\lambda\ll 1$ which can be regarded as a fine-tuning.  Hence, for the rest of this paper, we focus on flat directions that are lifted at the $n=4$ level (i.e. the smallest possible value of $n$). For this particular case, the condition $g^2|\varphi_{min}|^2<H_I^2$ translates to 
\be \label{E1}
g^2 \, < \, H_I/M_p \, , 
\ee
where $H_I$ is the Hubble parameter during inflation.

After inflation, the flat direction closely tracks the position of the minimum as it shifts to lower scales because of the time-dependence of the Hubble parameter. Once $c_HH^2\sim m_{\varphi}^2$, the flat direction gets a positive mass squared and starts oscillating with frequency $m_{\varphi}$. Since we are taking A-terms to be negligible, these oscillations are for all practical purposes one dimensional. They lead to a time-dependence of the $\chi$ field mass that can trigger non-perturbative amplification of $\chi$ fluctuations whenever the change of the $\chi$ frequency $\omega_{\chi}=g|\varphi|$ ceases to be adiabatic \cite{Kofman:1997yn}. We take the universe to be dominated by radiation after inflation. In a radiation dominated universe, the time evolution of the flat direction once it starts oscillating is given by:
\be\label{varphi_evo}
\varphi(t) \, = \, \frac{\varphi_0\sin{(m_{\varphi}t)}}{(m_{\varphi}t)^{3/4}}\equiv\Phi(t)\sin{(m_{\varphi}t)},
\ee
where $\varphi_0$ is given by (\ref{varphi_inf}) evaluated at $H(t)\simeq m_{\varphi}$. 

On the other hand, the perturbations in the $\chi$ field obey the wave equation in a FRW universe with metric $ds^2=dt^2-a^2(t)dx^2$:
\be
\ddot{\chi}_k + 3H\dot{\chi}_k + \left(\frac{k^2}{a^2}+g^2|\varphi(t)|^2\right)\chi_k \, = \, 0,
\ee
where $a$ is the usual scale factor. We can simplify this equation by rescaling the field as $X_k(t)=a^{3/2}(t)\chi_k(t)$ \cite{Kofman:1997yn}. This leads to the much simpler equation:
\be\label{eom_main1}
\ddot{X}_k+\omega_k^2X_k \, = \, 0,
\ee
where
\be\label{freq}
\omega_k^2\, = \, \frac{k^2}{a^2(t)}+g^2|\varphi(t)|^2-\frac{3}{4}H^2-\frac{3}{2}\frac{\ddot{a}}{a}.
\ee
In a radiation dominated universe, the two last terms of (\ref{freq}) add up to $+3H^2/4$ and thus $\omega_k^2$ is strictly positive. The oscillations of the flat direction may trigger the amplification of quantum fluctuations in $\chi$ via stochastic resonance.  Such a resonance will occur if the condition:
\be 
q_0 \, \equiv \, \frac{g^2\varphi_0^2}{4m_{\varphi}^2} \, = \, \frac{\sqrt{3c_H}g^2M_p}{2|\lambda|m_{\varphi}} \, \gg\, 1
\ee
is satisfied. Since we expect $c_H\sim\mathcal{O}(1)$ (see above) and taking $|\lambda|\sim\mathcal{O}(1)$ (which yields a lower bound on $q$), this condition is realized provided that 
\be \label{E2}
g^2M_p \, \gg \, m_{\varphi}
\ee
at the beginning of the resonance.  Assuming that this condition holds, the solution to (\ref{eom_main1}) is given by \cite{Kofman:1997yn}:
\be\label{sol_chi1}
|\chi_k(t)| \, \simeq \, \frac{|\chi_0(k)|e^{\mu_k m_{\varphi}t}}{a^{3/2}(t)},
\ee
where $\chi_0(k)$ is the initial amplitude of the $k^{th}$ mode of the $\chi$ fluctuations at the beginning of the resonance and $\mu_k$ is the effective Floquet exponent (whose value is discussed later). This growth corresponds to the squeezing of the quantum vacuum state of the $\chi$ field. The squeezed vacuum state is highly excited and
can be viewed as containing a condensate of $\chi$ particles with
momentum $k$. The solution (\ref{sol_chi1}) is only valid for modes satisfying: 
\be\label{resonance_band}
\frac{k^2}{a^2(t)} \, \lesssim \, \frac{gm_{\varphi}\Phi(t)}{\pi}.
\ee
For modes outside this resonance band, the adiabaticity condition is not violated. Hence, the effective Floquet exponent vanishes and no growth occurs. Note that large scale modes are preferably amplified compared to short wavelength modes. 

The resonant amplification takes place until backreaction shuts off the resonance. This occurs when the mass of the oscillating flat direction begins to be dominated by the fluctuations of the $\chi$ field, that is, when $g^2\langle\chi^2\rangle\simeq m_{\varphi}^2$. 

To summarize the results of this section, we have shown that
the oscillating homogeneous $\varphi$ field leads to stochastic
resonance of the $\chi$ field in the same way that the background
homogeneous inflation field leads to stochastic resonance of
matter fields which it couples to \cite{Kofman:1997yn,TB}, generating a squeezed
state of low frequency $\chi$ particles. The resonance of $\chi$
eventually shuts off the resonance by changing the time evolution
of the background $\varphi$ field.

\section{Change in the Curvature Perturbation on Large Scales}

We now turn our attention to the effect of the amplification of the $\chi$ fluctuations on large scale curvature perturbations. In the presence of a non-adiabatic pressure perturbation $\de p_{\text{nad}}$, the curvature perturbation on uniform-density hypersurfaces $\zeta$ varies according to the equation \cite{Lyth:1998xn,GarciaBellido:1995qq}:
\be
\dd{\zeta}\, =-\, \frac{H}{\rho+p}\de p_{\text{nad}},
\ee
where the non-adiabatic pressure perturbation is defined as:
\be
\de p_{\text{nad}} \, = \, \de p-\frac{\dd{p}}{\dd{\rho}}\de\rho.
\ee

In our case, the entropy mode is generated by the field $\chi$ and is
quadratic in $\chi$ (there is no rolling homogeneous $\chi$ background).
The leading-order pressure and energy density perturbations are given by:
\ba
\de\rho \, &=& \, \frac{1}{2}g^2\varphi^2\chi^2+\frac{1}{2}\dd{\chi}^2+\de\rho_a\\
\de p \, &=& \, -\frac{1}{2}g^2\varphi^2\chi^2+\frac{1}{2}\dd{\chi}^2+\de p_a,
\ea
where we defined $\varphi\equiv|\varphi|$ (we are considering radial motion of the field
only). $\de\rho_a$ and $\de p_a$ are the adiabatic component of the perturbations, that is, $\de p_a=(\dd{p_0}/\dd{\rho_0})\de\rho_a$. The adiabatic component of the fluid is made up of the field $\varphi$ and
of the radiation produced during inflationary reheating. Thus, the background 
density and pressure are given by:
\ba
\rho \, &=& \, \dd{\varphi}^2+m^2_{\varphi}\varphi^2+\rho_0\\
p \, &=& \, \dd{\varphi}^2-m^2_{\varphi}\varphi^2+p_0,
\ea
where $p_0$ and $\rho_0$ are the background pressure and energy density, subtracting the
contribution from $\varphi$. Since we are 
assuming that the flat direction decays during the post-inflationary phase of radiation 
domination, we have $p_0=(1/3)\rho_0$.  Combining the above results, we obtain 
the following expression for the non-adiabatic pressure perturbation:
\begin{widetext}

\be\label{non_adia_press}
\de p_{\text{nad}}
\, = \, \frac{\dd{\varphi}(g^2\chi^2\varphi^2\ddot{\varphi}-m_{\varphi}^2\dd{\chi}^2\varphi)+\left(\frac{4H\rho_0}{3}\right)(\frac{1}{2}\dd{\chi}^2-g^2\chi^2\varphi^2)}{3H\dd{\varphi}^2+2H\rho_0}
 -\frac{2\dd{\varphi}\de\rho_a}{3}\frac{2m_{\varphi}^2\varphi-\ddot{\varphi}}{3H\dd{\varphi}^2+2H\rho_0}.
\ee

\end{widetext}
Since we are interested in the growth of the non-adiabatic pressure perturbation at 
the time when the modulus field starts oscillating, we have $H\simeq m_{\varphi}$ such that 
$\rho_0\sim m_{\varphi}^2M_p^2$. Meanwhile, using (\ref{varphi_inf}), the time-averaged 
time derivative of the modulus field is $\langle\dd{\varphi}^2\rangle\sim m_{\varphi}^3M_p$. 
Therefore,
\be
\frac{\langle\dd{\varphi}^2\rangle}{\rho_0} \, \sim \, \frac{m_{\varphi}}{M_p} \, \ll \, 1,
\ee
and we can safely neglect the first term in the denominator of (\ref{non_adia_press}). Similarly, 
the second term of the numerator dominates over the first one since it is proportional to $\rho_0$.
Moreover, the last term of (\ref{non_adia_press}) is independent of $\chi$ (the field whose 
variance is growing exponentially) and we thus expect it to give a negligible contribution to 
the non-adiabatic pressure. Keeping only the dominant part to the non-adiabatic pressure 
perturbation, the equation of motion for the curvature perturbation reads:
\be
\dd{\zeta} \, \simeq \, -\frac{1}{6HM_p^2} (\frac{1}{2}\dd{\chi}^2-g^2\chi^2\varphi^2),
\ee
where we used $\rho_0=3M_p^2H^2$. Using $|\dd{\chi}|^2\simeq g^2\varphi^2|\chi|^2$ 
and $H=1/2t$, this equation can readily be integrated:
\be\label{zeta_nad_int}
\zeta_{\text{nad}} \, = \, \frac{g^2}{6M_pm_{\varphi}}|\chi_0|^2\int_{x_0}^{x_f}\frac{\sin^2{x}}{\sqrt{x}}e^{2\mu_kx}dx,
\ee
where $x=m_{\varphi}t$ and where we used (\ref{varphi_evo}) and (\ref{sol_chi1}) to obtain the 
time dependence of the fields.  The integral runs over the duration of the resonance  and the value 
of $ x_f$ corresponds to the time when the backreaction of the produced $\chi$ particles on the 
oscillating modulus becomes important, that is, when $g^2\langle\chi^2(x_f)\rangle\simeq m^2_{\varphi}$. 
Given the value of the Floquet exponent and the duration of the resonance, the integral in 
(\ref{zeta_nad_int}) can be done numerically and yields a pure number that we shall denote by $I$.\\

The power spectrum of the non-adiabatic curvature perturbations is given by:
\be
\mathcal{P}_{\zeta_{nad}} \, = \, \frac{k^3}{2\pi^2}|\zeta_{nad}(k)|^2,
\ee
where $\zeta_{nad}(k)$ is related to $\zeta_{nad}$ by a Fourier transform. Using (\ref{zeta_nad_int}) 
and performing the Fourier transform integral yields:
\be\label{non_PP}
\mathcal{P}_{\zeta_{nad}} \, = \, \frac{k^3}{4\pi}\left(\frac{g^2I}{6M_pm_{\varphi}}\right)^2\int d^3\kkk \frac{\mathcal{P}_{\chi}^{res}(|\kkk|)\mathcal{P}_{\chi}^{res}(|\kk-\kkk|)}{|\kkk|^3|\kk-\kkk|^3},
\ee
where $\mathcal{P}_{\chi}^{res}$ is the power spectrum of the $\chi$ particles at the beginning 
of the resonance which is given by:
\be
\mathcal{P}_{\chi}^{res} \, \simeq \, \left(\frac{H_I}{2\pi}\right)^2\left(\frac{m_{\varphi}}{H_I}\right)\left(\frac{k}{k_{end}}\right)^{2\de},
\ee
where $k_{end}$ is the last comoving scale to have exited the Hubble radius at the end of inflation
and $\de=g^2\varphi_I^2/3H_I^2$, where the $I$ subscript refers to a quantity evaluated during the 
inflationary era. The first term is term is the usual scale invariant spectrum for a field that is light during 
inflation, the second one accounts for the Hubble damping between the end of inflation and the beginning 
of the resonance while the k-dependent term takes into account the small mass of the $\chi$ field during 
inflation. Since we have $\de\ll1$, we can neglect the slight scale dependence of the power spectrum and 
perform the momentum integral. This integral has a IR divergence that must be regulated. A natural infrared cutoff is provided by the first comoving mode that exited the Hubble scale 
at the beginning of inflation.  Taking this mode to correspond to our present Hubble radius, we are basically integrating over today's visible universe, i.e. the only patch of the universe that we have information about.  The resonance band (\ref{resonance_band}) naturally provides a UV cutoff for the non-adiabatic power spectrum (although the answer depends very mildly on this choice since the integrand in (\ref{non_PP}) is rapidly converging). The answer then depends 
logarithmically on the total number of e-foldings of inflation. Taking $N=60$ and imposing the constraint 
$g^2<H_I/M_p$, we obtain an upper bound on the non-adiabatic power spectrum:
\ba
\mathcal{P}_{\zeta_{nad}}&\simeq&\frac{10^2}{2\pi^2}\left(\frac{g^2I}{6 M_p}\right)^2\left(\frac{H_I}{2\pi}\right)^2\en
&\lesssim&\frac{I^2}{10^2}\left(\frac{H_I}{M_p}\right)^4.
\ea
Given an inflationary scale $H_I$, the overall size of the non-adiabatic power spectrum is thus determined 
by the enhancement factor due to the resonant decay of the modulus field. Therefore, in order to see if this 
power spectrum could exceed the bound set by the COBE normalization, we need to obtain as estimate of 
the value of $I$ which in turn depends on the length of the resonance and on the value of the effective 
Floquet exponent.


\section{Strength of the Resonance} 

To estimate the resonance enhancement factor, one needs to determine the value of the effective 
Floquet exponent. On large scales, we can neglect the scale dependence of $\mu_k$, and its 
approximate value is then given by \cite{Kofman:1997yn}:
\be
\mu_k \, \equiv \, \mu \, \simeq \, \frac{1}{2\pi}\ln{3} \, \simeq \, 0.17.
\ee
The length of the resonance is obtained by solving the equation
$g^2\langle\chi^2(x_f)\rangle \, \simeq \, m_{\varphi}^2$
for $x_f$. The variance for the $\chi$ field at the end of the resonance is \cite{Linde:1982uu}:
\be
\langle\chi^2(x_f)\rangle \, \simeq \, \frac{3}{8\pi^2}\frac{H_I^4}{g^2\varphi_I^2}\frac{m_{\varphi}}{H_I}\left(\frac{a(x_0)}{a(x_f)}\right)^{3/2}e^{2\mu (x_f-x_0)}.
\ee
Using (\ref{varphi_inf}) to calculate the amplitude $\varphi$ during inflation and noting that 
$a(t)\propto t^{1/2}$, we can obtain an approximate solution (up  to a logarithmic correction) for $x_f$:
\be
x_f \, \simeq \, -\frac{1}{2\mu}\ln{\left[\left(\frac{H_I}{2\pi}\right)^2\frac{1}{M_pm_{\varphi}}\right]}+\mathcal{O}(\ln{x_f})
\ee
which is valid if $x_f\gg1$, that is, if $m_{\varphi}\gg H_I^2/M_p$, which is also the condition that 
must be satisfied to have a long-lasting resonance. Note that this result is independent of the coupling 
constant $g$. 

\begin{widetext}
\begin{figure}[t!]
\includegraphics[width=\textwidth]{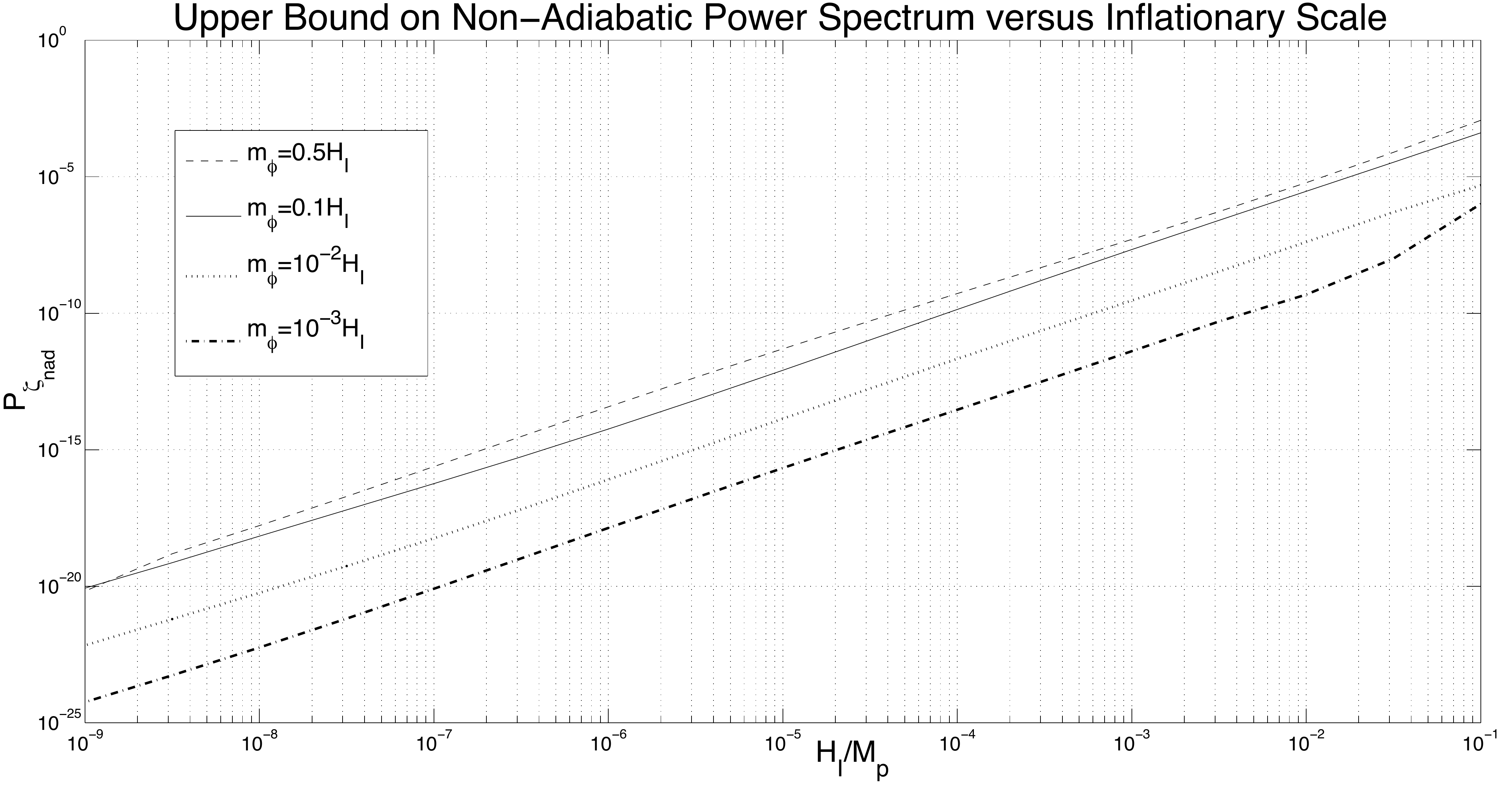}
\caption{Upper bound for the non-adiabatic power spectrum for different values of the modulus mass. For $m_{\varphi}\simeq10^{-1}H_I$, the resonance could lead to an observable effect only if $H_I\gtrsim10^{-4}M_p$ while for  $m_{\varphi}\simeq10^{-3}H_I$, one needs $H_I\gtrsim10^{-2}M_p$. \label{result_graph}}
\end{figure}
\end{widetext}

Combining this result with the requirement that the mass of $\varphi$ is negligible during
the period of inflation and with the constraints on $g$ found earlier (see (\ref{E1} and \ref{E2}), we 
obtain the following parameter space for which resonance could possibly lead to a curvature power 
spectrum that exceeds current observational bounds:
\ba
H_I \, > \, m_{\varphi} \, \gg \, \frac{H_I^2}{M_p}\\
\frac{m_{\varphi}}{M_p} \, \ll \, g^2 \, < \, \frac{H_I}{M_p}.
\ea
The crucial question now is: given values of parameters lying within the above bounds, is it possible for 
the resonance to lead to an observable effect on the curvature power spectrum? First, to obtain an upper 
bound on the duration of the resonance, we take $m_{\varphi}$ to be close to the inflation scale, that is, 
we write $m_{\varphi}=\epsilon H_I$. We can now integrate (\ref{zeta_nad_int}) and calculate the upper 
bound on the non-adiabatic curvature power spectrum as a function of the inflationary scale. The results 
are shown in Figure \ref{result_graph} for various values of $\epsilon$.

We see that the non-adiabatic power spectrum can exceed the limit set by COBE \cite{Bunn:1996py} 
only if the mass of the modulus field is close to the inflationary scale and if the latter lies close to the Planck 
scale. For example, if $m_{\varphi}\simeq10^{-3}H_I$, then the non-adiabatic power spectrum could exceed 
observational bound only if $H_I\gtrsim10^{-2}M_p$. Lower moduli masses or inflationary scales yield 
non-adiabatic power spectra that do not contradict observational bound and therefore do not allow us to 
put constraints on the particle physics parameters.  For moduli masses larger than $10^{-3}H_I$, there is a 
region of the $(m,g)$ parameter space which is excluded, provided that the inflationary scale is larger 
than $10^{-4}M_p$. For example, for $H_I\simeq10^{-3}M_p$ the following values could be excluded:
\ba
10^{-3} \, > \, \frac{m_{\varphi}}{M_p} \, \gg \, 10^{-6}\\
\left(\frac{m_{\varphi}}{M_p}\right)^{1/2} \, \ll \, g \, < \, 10^{-3/2}.
\ea
In particular, the usual value $m_{\varphi}\sim\mathcal{O}(\text{TeV})$, which allows SUSY to solve the 
hierarchy problem, is permitted for all values of the inflationary scale.\\


\section{Discussion and Conclusion}

We have shown that for a MSSM modulus field which acquires a negative Hubble-induced mass 
during inflation and subsequently decays via broad resonance, the growth of the curvature perturbations 
due to the non-adiabatic pressure does not contradict current observational bound for most of the parameter 
space. The non-perturbative
decay of the modulus field could yield an observable effect only if the scale of inflation is very close to the 
Planck scale and if the soft SUSY-breaking mass is close to the inflationary scale. Consequently, the technique 
developed above does not allow one to put constraints on the MSSM at the TeV scale. However, this result 
depends closely on the shape of the modulus potential (\ref{pot_final}). In particular, the effect can only be 
large if the energy stored in the moduli field is a considerable fraction of the overall energy density of the universe. 

In this paper, we treated (inaccurately) the modulus field as a real field. However, MSSM flat 
directions are intrinsically complex fields and one could ask if our result apply if we consider the complex 
nature of the moduli. In references \cite{Olive:2006uw,Gumrukcuoglu:2008fk,Basboll:2007vt,Basboll:2008gc}, 
it was shown that by taking into account their complex nature, flat directions can decay non-perturbatively if 
at least two flat directions are excited at the same time. Therefore, it seems reasonable to expect that 
parametric amplification of entropy modes (and by consequence of the curvature perturbations) can happen 
for more realistic model parameter values if we were to consider the full complex nature of flat directions.

In conclusion, we point out that the large scale curvature perturbations that were amplified by stochastic 
resonance could be the dominant contribution to the curvature power spectrum. Indeed, within the 
model developed above, inflation does not have to generate perturbations of the right order of magnitude. 
Instead,  these could be generated at later times via the stochastic decay of a modulus field. 

\acknowledgments

We thank Jean Lachapelle, Aaron C. Vincent, Nima Lashkari, Toni Riotto, Omid Saremi, James Sully and Kris Sigurdson for useful discussion. FYCR is supported by NSERC. RB wishes
to thank the Theory Division of the Institute of High Energy
Physics (IHEP) for their wonderful hospitality and financial support.
RB is also supported by an NSERC Discovery Grant and 
by the Canada Research Chairs Program. 

\bibliography{Mssm_resonance}

\end{document}